# Spin Glass Behavior in the $Dy_{3-x}Y_xTaO_7$ (0 ≤ x ≤ 1) System

J. Francisco Gomez-Garcia[1], Roberto Escudero[2] and Gustavo Tavizon[1]*

[1]Departamento de Física y Química Teórica, Facultad de Química, Universidad Nacional Autónoma de México, Av. Universidad 3000, Coyoacán, México D. F. 04510. México.

[2]Departamento de Materia Condensada y Criogenia, Instituto de Investigaciones en Materiales, Universidad Nacional Autónoma de México, Av. Universidad 3000, Coyoacán, México D. F. 04510. México.

*Corresponding author: Phone: 52(55) 56223831; Fax: 52(55) 56162010;
Email: gtavizon@unam.mx

**Abstract**

Several x-compositions of the polycrystalline $Dy_{3-x}Y_xTaO_7$ system, crystallizing in the weberite-type structure, were synthesized and structurally characterized using Rietveld refinements based on X-ray diffraction data. In previous magnetic characterization of $Dy_3TaO_7$ (x = 0), with the same crystal structure, an antiferromagnetic transition at T = 2.3 K has been assigned to this compound. On the basis of DC and AC magnetic susceptibilities analyses, we show in this work that all compounds in the range of 0 ≤ x ≤ 1.0 exhibit a spin glass behavior. The nature of the spin glass behavior in $Dy_{3-x}Y_xTaO_7$, can be attributed to the highly frustrated antiferromagnetic interaction of the $Dy^{3+}$ sublattice and to the $Dy^{3+}$-$Dy^{3+}$ distorted tetrahedra array in the weberite-type structure of this system. By fitting AC susceptibility data, using dynamical scaling theory equations, we conclude that a cluster spin glass is present in $Dy_{3-x}Y_xTaO_7$ in the low temperature range. Depending on the x-composition, $T_g$ ~ 2.2 - 3.2 K. In the range 15-300 K the system obeys a Curie-Weiss magnetic behavior.

## 1. Introduction

Complex oxides with nominal formula $Ln_3MO_7$, where Ln is a trivalent lanthanide or yttrium and M is a pentavalent metal cation, exhibit a weberite related structure. These oxides attract great attention because they display interesting properties as dielectric materials [1,2], magnetic systems [3–9], and when M = Ta, as a possible electrolyte for solid oxide full cell [10], as well as being a possible heterogeneous photocatalyst for hydrogen generation from water splitting [11,12]. According to Alpress and Rossell [13,14], for M = Nb, Ta and Sb, three types of structures, depending on the lanthanide size, can be obtained: from a cubic fluorite-type structure to an orthorhombic $C222_1$. In fact, the non-cubic weberite-type crystal structures of the $Ln_3TaO_7$ compounds [14] can be described as an anion-deficient fluorite-related superstructure [15].

The non-centrosymmetric weberite-type compounds ($C222_1$) have structural characteristics that could reasonably yield peculiar magnetic and electronic properties. One of these characteristics is related to the $MO_6$ octahedra arrangement, which is quasi-one-dimensional; these are distorted corning-sharing octahedra with zig-zag chains parallel to the $c$-axis (see Fig. 1). Another aspect is given by the Ln-Ln sublattice; where the Ln cations form a complex 3-D array of corner and edge-sharing slightly distorted tetrahedra as can be observed in Fig. 1. If $M^{5+}$ is a non-magnetic ion, and $Ln^{3+}$ is magnetic, the distorted tetrahedra arrangement constitutes the magnetic lattice in the $Ln_3MO_7$ system.

Additionally, magnetic structures which combine both antiferromagnetism and lattice geometry, based on triangles and tetrahedra, inhibit the formation of a collinear ordered state and often display geometric magnetic frustration [16]. As pointed out by Fennel et al. for $Ho_3SbO_7$ crystallizing in the $C222_1$ SG [17], the Ln geometry sublattice (closely related to the pyrochlore lattice) is expected to be a candidate for highly frustrated magnetism. In our knowledge, there have been no reports on the $Ln_3MO_7$ system in which the Ln or M magnetic ions order ferromagnetically. Paramagnetic behavior in weberite-type systems has been reported in several works [3,7–9]; with antiferromagnetic (AFM) order [5,7,9,17]; with weak ferromagnetism [18]; and with ferrimagnetic ordering [19]. Finally two compounds have been pointed out as probable spin glass systems [3,6].

For the $Dy_3TaO_7$ system ($C222_1$ SG) Wakeshima et al. observed a broad peak at 2.8 K in the temperature (T) dependent magnetic susceptibility ($\chi$); they assigned this behavior to an antiferromagnetic ordering of the $Dy^{3+}$ ions [7]. In the present work, single phase polycrystalline compounds of the $Dy_{3-x}Y_xTaO_7$ solid solution (0 ≤ x ≤ 1.0) were synthesized. Through X-ray diffraction measurements and crystal structure refinements, their structures were carefully determined. The DC magnetic susceptibility was measured from 2 K to room temperature (RT). The temperature dependence of the AC magnetic susceptibility of samples was measured between 2 and 20 K at frequencies of 50, 250, 500, 750 and 1000 Hz. The AC driven field was 1 Oe, and no external DC magnetic field was applied.

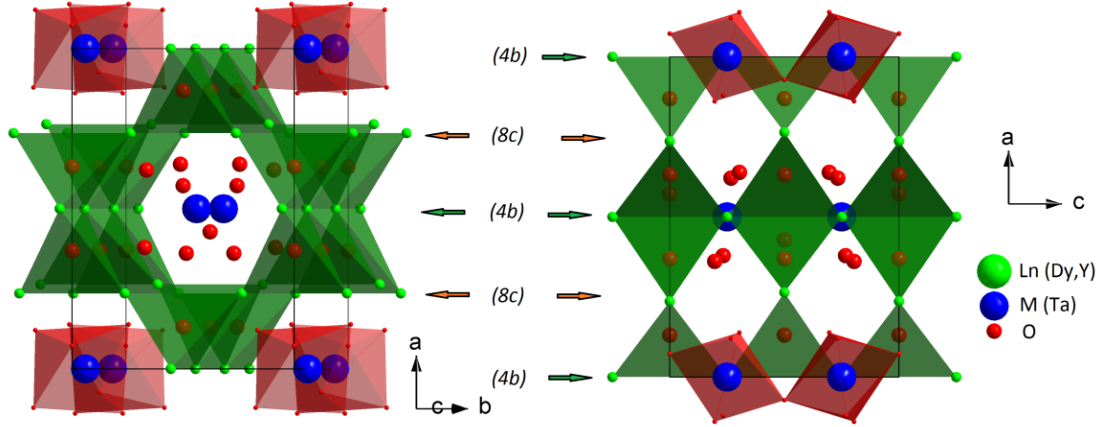

Fig. 1. (Color online) Crystal structure of the $Dy_3TaO_7$ weberite with the *C222$_1$* SG (No. 20). The $TaO_6$ octahedra (in red color) and the arrangement of the Dy-Dy distorted tetrahedra at the second-nearest neighbor site (in green color) are shown. The Wyckoff positions occupied by $Ln^{3+}$ are indicated in parenthesis between each image.

## 2. Material and Methods

Four compositions of the $Dy_{3-x}Y_xTaO_7$ system were synthesized by the conventional solid state reaction method. Stoichiometric amounts of $Dy_2O_3$, $Y_2O_3$ and $Ta_2O_5$ (Sigma-Aldrich, purity 99.99%) were weighted according to equation 1 with x = 0, 0.33, 0.66, and 1.0. The mixed stoichiometric powders were ground well in an agate mortar with acetone and then pressed at 350 MPa using a uniaxial press. The pellets obtained were calcined in alumina crucibles at 1400 °C in air for two days, with intermediate regrinding, and sintered at 1600 °C for six hours; white pellets with high hardness were obtained.

$$\left(\frac{3-x}{2}\right)Dy_2O_3 + \frac{x}{2}Y_2O_3 + \frac{1}{2}Ta_2O_5 \longrightarrow Dy_{3-x}Y_xTaO_7 \qquad \text{Equation 1}$$

Structural characterization was carried out by X-ray powder diffraction in a Bruker D8 Advance diffractometer with Cu $k_{\alpha 1}$ radiation (λ = 1.54184 Å) coupled with lynx eye detector. The patterns were collected in a 2θ range from 10° to 110°, with a 0.02° step/1.5 seconds at room temperature. Further Rietveld analysis was performed for each pattern using the GSAS code with the EXPGUI graphical interface [20,21]. Magnetic measurements were conducted in a quantum interference device (SQUID) magnetometer (MPMS, Quantum Design) coupled with an AC device. The DC magnetic measurements were performed at 100 Oe with temperature ranging from 2 to 300 K in zero-field-cooled (ZFC) and field-cooled (FC) modes. The AC measurements were done at 1 Oe with temperature ranging from 2 to 20 K, and a frequency ranging from 50 to 1000 Hz.

## 3. Results and discussion

### 3.1 X-ray diffraction

The powder X-ray diffraction (XRD) patterns at room temperature show that all synthesized samples are single-phase systems crystallizing in the $Ln_3MO_7$ (Ln = Dy and Y, and M = Ta) [22] weberite-type structure with no detectable secondary phases. In order to determine accurate lattice parameters and atomic positions, Rietveld structural refinements were performed for all the powder XRD data. All samples were successfully indexed by an orthorhombic lattice with space group *C222$_1$* (SG No. 20) [7]. Fig. 2 shows the experimental and calculated XRD patterns of samples with x = 0, 0.33, 0.66 and 1.0 respectively. The low $\chi^2$ and $R_{wp}$ values (shown in the plots) indicate a very good goodness-of-fit of the model to the experimental diffraction data.

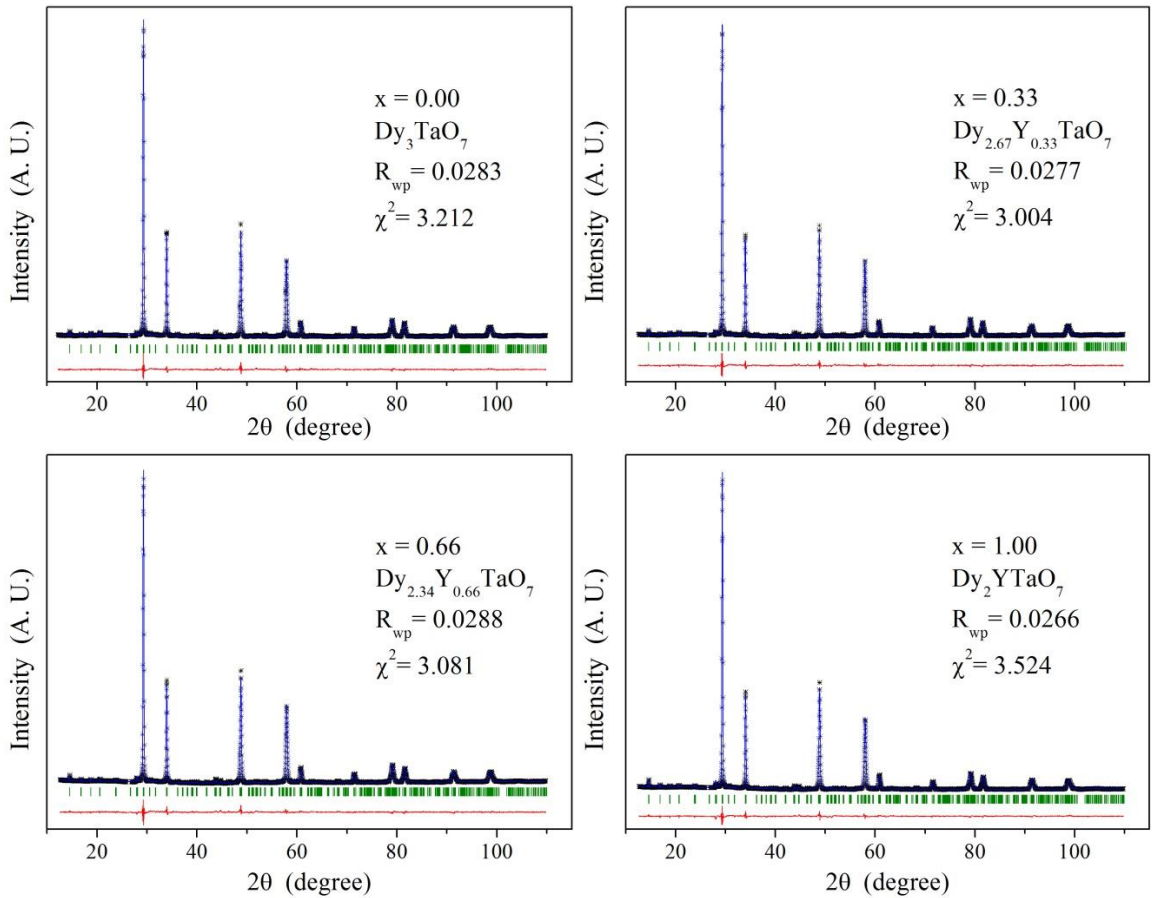

Fig. 2. (Color online) XRD patterns of $Dy_{3-x}Y_xTaO_7$ samples with x = 0, 0.33, 0.66, and 1.0. Blue line shows the calculated pattern and the experimental pattern data are indicated by black crosses; the difference between the experimental and calculated patterns is in red lines at the bottom of each plot; the green bars represent the Bragg-peak positions. The goodness-of-fit parameters are also indicated.

Due to the significant differences in the radii of the M ($Ta^{5+}$) and Ln ($Dy^{3+}$, $Y^{3+}$) cations in the weberite-type structure ($Ta^{5+}$, $Y^{3+}$ and $Dy^{3+}$ ionic radii are 0.64, 0.96 and 0.97 Å, respectively, all in coordination number (CN) 7 [23]), and according to the rule of parsimony for ionic crystals [24], an ordering of cations can be assumed in which all the $Y^{+3}$ substitutions occur at the $Dy^{3+}$ sites (excluding the $Ta^{5+}$ (*4b*) site). On the other hand, lattice parameters for $Dy_3TaO_7$ (x = 0) are almost the same as those reported by Wakeshima [7], and the change in the unit-cell volume for the composition range x = 0 - 1.0 is about 0.5 %, which is consistent with the similitude between the $Y^{3+}$ and $Dy^{3+}$ ionic radii (see Table I-IV). In this way, even $Dy^{3+}$, at the *4b* Wyckoff position (WP), and $Dy^{3+}$ at the *8c* WP are distinguishable by the CN (eight and seven, respectively), there is not any reasonable assumption to assign one particular site to the $Dy^{3+}$ substitution by $Y^{3+}$.

Crystal cell parameters, refined atomic positions, thermal factors, and goodness-of-fit criteria of the Rietveld refinements are presented in Tables I-IV. As is well known, metallic cations used in this work show a notable stability in their oxidation numbers: Dy (III), Y (III) and Ta (V). On the other hand, taking into account the low volatility of the Ta, Dy and Y-oxides, no deviations in the oxygen stoichiometry of samples was assumed and the site occupation factors (SOF) were not fitted in the structure refinements, they were fixed to the stoichiometric values. The change in lattice parameters as a function of The $Dy^{3+}$ content is shown in Fig. 3. According to the small difference between the $Dy^{3+}$ and $Y^{3+}$ ionic radii, a very small slope of the linear behavior is observed. The $Dy_3TaO_7$ unit cell is shown in Fig. 1.

*Table I. Atomic positions and isotropic thermal parameters for $Dy_3TaO_7$.*

| Atom | Wyckoff Position | $Dy_3TaO_7$, x = 0 | | | | |
|---|---|---|---|---|---|---|
| | | S. O. F. | x/a | y/b | z/c | U (Å$^2$) |
| Dy1 | 4b | 1 | 0 | 0.4956(8) | ¼ | 0.25(5) |
| Dy2 | 8c | 1 | 0.2376(2) | 0.2542(9) | 0.0026(8) | 0.001(1) |
| Ta | 4b | 1 | 0 | 0.0174(7) | ¼ | 0.023(6) |
| O1 | 8c | 1 | 0.104(2) | 0.242(5) | 0.302(4) | 0.031(2) |
| O2 | 8c | 1 | 0.129(2) | 0.805(3) | 0.261(5) | 0.025(4) |
| O3 | 4a | 1 | 0.143(3) | ½ | 0 | 0.031(7) |
| O4 | 4a | 1 | 0.141(3) | ½ | ½ | 0.006(4) |
| O5 | 4a | 1 | 0.100(3) | 0 | 0 | 0.013(3) |

*a*=10.5453(2), *b*=7.4583(1), and *c*=7.4963(1) Å; V= 589.58(2) Å$^3$; $R_{WP}$= 0.0283; $\chi^2$= 3.212

*Table II. Atomic positions and isotropic thermal parameters for $Dy_{2.67}Y_{0.33}TaO_7$.*

| Atom | Wyckoff Position | $Dy_{2.67}Y_{0.33}TaO_7$, x = 0.33 | | | | |
|---|---|---|---|---|---|---|
| | | S. O. F. | x/a | y/b | z/c | U (Å$^2$) |
| Dy1 | 4b | 0.89 | 0 | 0.4965(6) | ¼ | 0.016(6) |
| Y1 | 4b | 0.11 | 0 | 0.4965(6) | ¼ | 0.016(6) |
| Dy2 | 8c | 0.89 | 0.2376(2) | 0.2505(9) | 0.000(1) | 0.013(5) |
| Y2 | 8c | 0.11 | 0.2376(2) | 0.2505(9) | 0.000(1) | 0.013(5) |
| Ta | 4b | 1 | 0 | 0.0145(4) | ¼ | 0.002(1) |
| O1 | 8c | 1 | 0.104(2) | 0.264(4) | 0.300(4) | 0.030(2) |
| O2 | 8c | 1 | 0.136(2) | 0.798(3) | 0.265(5) | 0.003(3) |
| O3 | 4a | 1 | 0.141(4) | ½ | 0 | 0.050(8) |
| O4 | 4a | 1 | 0.141(3) | ½ | ½ | 0.018(3) |
| O5 | 4a | 1 | 0.104(4) | 0 | 0 | 0.022(5) |

$a$=10.5407(2), $b$=7.4560(1), and $c$=7.4921(1) Å; V=588.82(2) Å$^3$; R$_{WP}$=0.0277; $\chi^2$=3.004

*Table III. Atomic positions and isotropic thermal parameters for $Dy_{2.34}Y_{0.66}TaO_7$.*

| Atom | Wyckoff Position | $Dy_{2.34}Y_{0.66}TaO_7$, x = 0.66 | | | | |
|---|---|---|---|---|---|---|
| | | S. O. F. | x/a | y/b | z/c | U (Å$^2$) |
| Dy1 | 4b | 0.78 | 0 | 0.4973(7) | ¼ | 0.012(5) |
| Y1 | 4b | 0.22 | 0 | 0.4973(7) | ¼ | 0.012(5) |
| Dy2 | 8c | 0.78 | 0.2380(2) | 0.250(1) | 0.000(1) | 0.012(4) |
| Y2 | 8c | 0.22 | 0.2380(2) | 0.250(1) | 0.000(1) | 0.012(4) |
| Ta | 4b | 1 | 0 | 0.0139(5) | ¼ | 0.004(3) |
| O1 | 8c | 1 | 0.102(2) | 0.264(4) | 0.298(4) | 0.030(5) |
| O2 | 8c | 1 | 0.140(2) | 0.796(3) | 0.263(5) | 0.003(2) |
| O3 | 4a | 1 | 0.139(4) | ½ | 0 | 0.052(3) |
| O4 | 4a | 1 | 0.143(3) | ½ | ½ | 0.019(7) |
| O5 | 4a | 1 | 0.105(3) | 0 | 0 | 0.021(7) |

$a$=10.5353(2), $b$=7.4517(1), $c$= 7.4873(1) Å; V=587.80(2) Å$^3$; R$_{WP}$=0.0288; $\chi^2$=3.081

*Table IV. Atomic positions and isotropic thermal parameters for $Dy_2YTaO_7$.*

| Atom | Wyckoff Position | $Dy_2YTaO_7$, x = 1.00 | | | | |
|---|---|---|---|---|---|---|
| | | S. O. F. | x/a | y/b | z/c | U (Å$^2$) |
| Dy1 | 4b | 0.667 | 0 | 0.4983(7) | ¼ | 0.021(5) |
| Y1 | 4b | 0.333 | 0 | 0.4983(7) | ¼ | 0.021(5) |
| Dy2 | 8c | 0.667 | 0.2380(2) | 0.2469(9) | 0.000(1) | 0.016(2) |
| Y2 | 8c | 0.333 | 0.2380(2) | 0.2469(9) | 0.000(1) | 0.016(2) |
| Ta | 4b | 1 | 0 | 0.0138(4) | ¼ | 0.004(2) |
| O1 | 8c | 1 | 0.104(2) | 0.279(3) | 0.289(5) | 0.052(2) |
| O2 | 8c | 1 | 0.136(2) | 0.794(3) | 0.271(5) | 0.012(1) |
| O3 | 4a | 1 | 0.151(3) | ½ | 0 | 0.037(5) |
| O4 | 4a | 1 | 0.138(3) | ½ | ½ | 0.019(1) |
| O5 | 4a | 1 | 0.099(2) | 0 | 0 | 0.014(5) |

$a$=10.5316(2), $b$=7.4504(1), $c$=7.4840(1) Å; V=587.23(2) Å$^3$; R$_{WP}$=0.0266; $\chi^2$=3.524

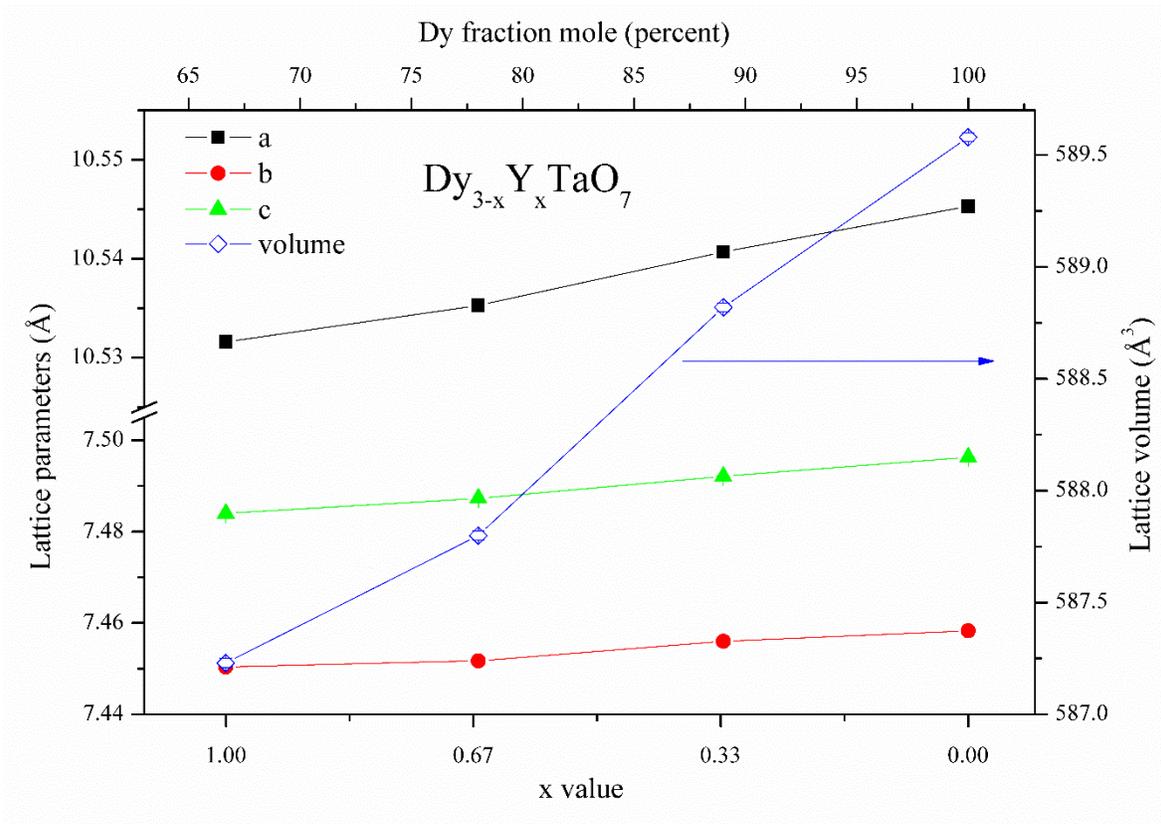

Fig. 3. (Color online) Lattice parameters as a function of the $Y^{3+}$ content. The linear behavior is associated with the formation of a substitutional solid solution in the $Dy_{3-x}Y_xTaO_7$ system.

The distortions in the $Dy^{3+}$-$Dy^{3+}$ tetrahedra can result from the low symmetry of *4b* and *8c* WP of $Ln^{3+}$ (Ln = Dy and Y) in $Dy_{3-x}Y_xTaO_7$. For x=0, $Dy^{3+}$ occupies a *4b* WP, with coordinates (*0, y, ¼*), and this cation is coordinated with other ten $Dy^{3+}$ ions in the second-nearest neighborhood (SNN), when Ta is omitted. In this WP, $Dy^{3+}$ shows five different $Dy^{3+}$-$Dy^{3+}$ lengths (3.600, 3.653, 3.748, 3.836 and 3.847 Å). Besides, the $Dy^{3+}$ at the *8c* WP, which is a general position (*x, y, z*), has only eight $Dy^{3+}$ ions in the SNN, with seven different $Dy^{3+}$-$Dy^{3+}$ length (3.600, 3.653, 3.666, 3.757, 3.792, 3.836 and 3.847 Å). As can be noted from the Table V data, there is not a clear relation between the $Y^{3+}$ content in $Dy_{3-x}Y_xTaO_7$ and the tetrahedra distortions, however, the average Ln-Ln length (see Table V) decreases as the $Y^{3+}$ content increases. This can be indicative of a random occupation of $Y^{3+}$ at *4b* and *8c* sites.

Table V. $Ln^{3+}$-$Ln^{3+}$ lengths (in Å) for the $Dy_{3-x}Y_xTaO_7$ system. (*) is used to indicate those that appear twice.

| $Ln^{3+}$ site in the SNN | $Ln^{3+}$ in 4b site | | | | $Ln^{3+}$ site in the SNN | $Ln^{3+}$ in 8c site | | | |
|---|---|---|---|---|---|---|---|---|---|
| | x = 0 | x = 0.33 | x = 0.66 | x = 1.0 | | x = 0 | x = 0.33 | x = 0.66 | x = 1.0 |
| 8c(*) | 3.600 | 3.626 | 3.640 | 3.618 | 4b | 3.600 | 3.626 | 3.640 | 3.618 |
| 8c(*) | 3.653 | 3.660 | 3.641 | 3.630 | 4b | 3.653 | 3.660 | 3.641 | 3.630 |
| 4b(*) | 3.748 | 3.746 | 3.743 | 3.742 | 8c | 3.666 | 3.713 | 3.691 | 3.656 |
| 8c(*) | 3.836 | 3.812 | 3.818 | 3.833 | 8c(*) | 3.757 | 3.742 | 3.750 | 3.750 |
| 8c(*) | 3.847 | 3.833 | 3.821 | 3.834 | 8c | 3.792 | 3.755 | 3.760 | 3.794 |
| Average | 3.737 | 3.735 | 3.733 | 3.731 | 4b | 3.836 | 3.812 | 3.818 | 3.833 |
| | | | | | 4b | 3.847 | 3.833 | 3.821 | 3.834 |
| | | | | | Average | 3.739 | 3.735 | 3.734 | 3.733 |

## 3.2 DC magnetization

The temperature dependence of magnetic susceptibility for x = 0, 0.33, 0.66, and 1.0, in the ZFC and FC modes under an applied field of 100 Oe is shown in the Fig. 4A. As depicted in this plot, a clear paramagnetic Curie-Weiss (CW) behavior for all samples is present above 15 K; below this temperature a broad magnetic signal, at about 2.2 K, is present in samples with x = 0, 0.33 and 0.66. The magnetic behavior of $Dy_3TaO_7$, (x = 0), in this temperature range, has previously been assigned to an AFM transition [7]. The data above 15 K, after subtracting the diamagnetic contribution of the cores [25], can be well fitted by the CW law yielding Curie-Weiss temperatures ($\Theta_{CW}$) of -11.2, -16.8, -12.8, and -9.7 K, and the Curie constants (C) were 40.0, 37.2, 32.6, and 26.1 emu/mol K, for x = 0, 0.33, 0.66, and 1.0, respectively. The negative values of $\Theta_{CW}$ are indicative of a moderate antiferromagnetic coupling between the $Dy^{3+}$ ions of the magnetic lattice. The values found for $\Theta_{CW}$ in $Dy_{3-x}Y_xTaO_7$ are in the order of those exhibited by rare earth titanate pyrochlores in which a clear geometric magnetic frustration has been reported [26]. From the $\chi^{-1}$ vs. temperature plots, the estimated effective magnetic moments for $Dy^{3+}$ are 10.33 (x = 0), 10.56 (x = 0.33), 10.55 (x = 0.66), and 10.24 (x = 1.0), all in units of Bohr magnetons (BM). These values are in good agreement with those obtained from the Russell-Saunders coupling of spin and orbital angular momenta for isolated $Dy^{3+}$ ($4f^9$, $^6H_{15/2}$) 10.63 BM [27]. In Table VI are shown the main magnetic parameters and the frustration factors, $f_f = |\theta_{CW}|/T_f$, the values of the latter indicates that magnetic frustration is present in the system [28].

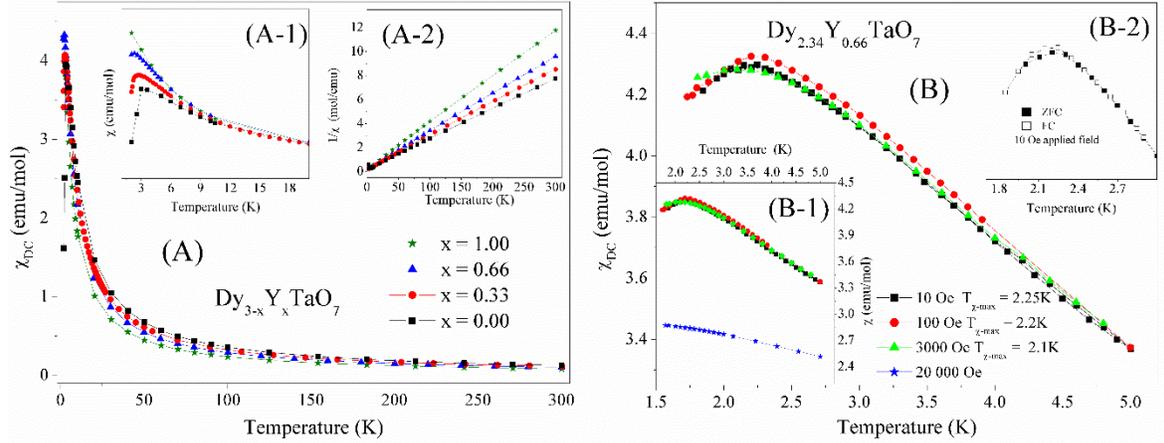

Fig. 4. (Color online) DC magnetic measurements. (A) ZFC magnetic susceptibility of the $D_{3-x}Y_xTaO_7$ samples; $\mu_0H$= 100 Oe. The inset A-1 shows the low temperature region and the maximum in each curve; (A-2) corresponds to a $1/\chi$ vs. T plot for the high temperature region. (B) Temperature dependence of magnetic susceptibility for the $Dy_{2.34}Y_{0.66}TaO_7$ sample at several DC magnetic fields. In (B-1) the maximum at ~2.2 K disappears in a strong magnetic field (20 kOe). In (B-2), divergence in the ZFC and FC susceptibilities for x = 0.66 under a DC magnetic field of 10 Oe.

*Table VI. Curie-Weiss parameters obtained from linear fit of $1/\chi$ against T curves. The frustration factor $f_f = |\theta_{CW}|/T_f$, is also presented.*

| Sample | x value | C (emu/mol K) | $\theta_{CW}$ (K) | $T_{\chi\text{-max}}$ (K) | $\mu_{eff}$ ($\mu_B$) | $f_f$ |
|---|---|---|---|---|---|---|
| $Dy_3TaO_7$ | 0 | 40 | -11.2 | 3 | 10.33 | 3.73 |
| $Dy_{2.67}Y_{0.33}TaO_7$ | 0.33 | 37.2 | -16.8 | 2.7 | 10.56 | 6.22 |
| $Dy_{2.34}Y_{0.66}TaO_7$ | 0.66 | 32.6 | -12.8 | 2.3 | 10.55 | 4.57 |
| $Dy_2YTaO_7$ | 1.00 | 26.1 | -9.7 | -- | 10.24 | >4.85* |

*no maximum in DC magnetization was observed, and $T_f$ was estimated from AC measurements.

Since in the sample with x = 0.66, $Dy_{2.34}Y_{0.66}TaO_7$, a broad peak around 2.25 K was also present, additional DC magnetization measurements under magnetic fields ($\mu_0H$) of 10, 100, 3000, and 20 000 Oe were performed in the ZFC and FC modes; Fig. 4B accounts for such results. As can be observed in the inset B1, as the probing field H increases, the peak becomes wider, and the humps show a slight shift to a lower temperature with increasing magnetic field. Moreover, the cusp of the magnetic susceptibility almost disappears at 20 000 Oe; this magnetic behavior has been previously observed in another spin glass system as $BaCo_6Ti_6O_{19}$ [29]. As is well known, this temperature dependence of the transition temperature on the probe magnetic field (see Fig. 4B-1) is not the typical behavior expected for an AFM ordering, even less if the obtained negative values of $\Theta_{CW}$ are associated with a moderate AFM coupling. As an additional observation from this plot, the different magnetization values for the ZFC and FC modes are only present for the lowest value of the probing field H (10 Oe); probably, for H > 10 Oe, the probing field perturbs magnetic interactions in the $Dy_{2.34}Y_{0.66}TaO_7$ system. Grossly, the irreversibility temperature (IT) for $Dy_{2.34}Y_{0.66}TaO_7$ under $\mu_0H$ = 10 Oe should be around 2.25 K. In the isothermal magnetization of $Dy_{2.34}Y_{0.66}TaO_7$, depicted in Fig. 5 for T = 2 and 10 K, no hysteretic behavior can be observed and the magnetic saturation at 2 K is almost reached at $\mu_0H$ = 30 000 Oe. The lack of remnant magnetization in $Dy_{2.34}Y_{0.66}TaO_7$ implies that there is not a net magnetization as those associated to ferromagnetism (FM) or to weak ferromagnetism (WFM) in these samples. It is worth mentioning here that the isothermal magnetization was followed at 2 K, and this temperature is on the edge ($T_g$ = 2.3 K, see below) of the paramagnetic state of this system. On the other hand, in the absence of anisotropy, the field-cooled and the zero-field-cooled magnetization are macroscopically equivalent and magnetic hysteresis is absent [30]. In terms of the $Dy^{3+}$ content of samples, the maxima in the $\chi$ vs. T curves ($T_{\chi-max}$) are at 2.3 K for $Dy_{2.34}Y_{0.66}TaO_7$, 2.7 K for $Dy_{2.66}Y_{0.34}TaO_7$, 3.0 K for $Dy_3TaO_7$, and no hump is observed in the DC magnetization of $Dy_2YTaO_7$.

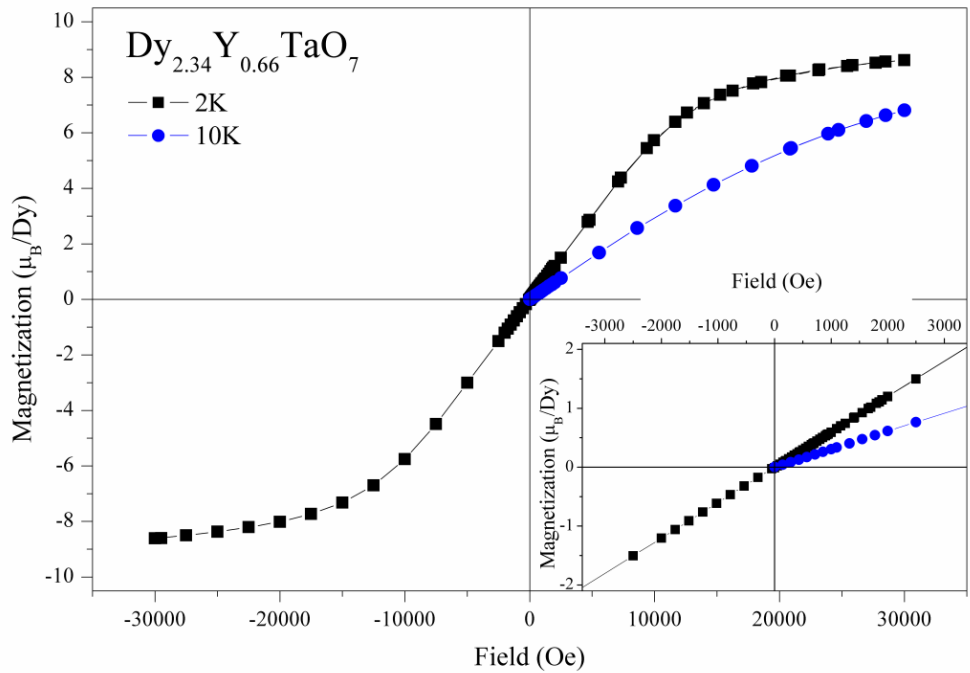

Fig. 5. Isothermal magnetization of $Dy_{2.34}Y_{0.66}TaO_7$ at 2 and 10 K. No saturation is observed at 10 K, while the 2 K magnetization shows that saturation is almost reached at $\mu_0H$ = 30 000 Oe. The inset shows the low field range. As can be noted, no hysteretic effect is observed and the sample behaves as a paramagnet, even for T = 2 K (For this sample $T_g$ = 2.21 K, see below).

### 3.3 AC magnetic susceptibility

DC magnetization experiments cannot conclusively identify a spin glass. Spin glass behavior is usually studied by AC susceptibility measurements, and the spin glass transition temperature is accurately determined by the frequency dependence of real ($\chi'$) or imaginary ($\chi''$) components of the AC susceptibility [16,31]. In order to further investigate the nature of the broad peak around 2-3 K in the DC magnetization of $Dy_{3-x}Y_xTaO_7$ (x = 0, 0.33 and 0.66), AC magnetic measurements were performed on samples with these compositions (x = 1.0 was also included). The temperature dependence of the $\chi'$ and $\chi''$ components at different frequencies are plotted in Fig. 6. From the $\chi'$ vs. T plots the temperature $T_f$ (maximum AC susceptibility temperature) at the maximum of this broad peak shifts to higher values as the frequency increases, as can be observed for all x-compositions in the $0 \leq x \leq 1$ range. This is a characteristic behavior of a spin glass compound [31]. The temperature dependence of the imaginary components $\chi''(T)$ (see Fig. 5) show the corresponding shifts of $T_f$, and the magnitude of the peak in $\chi''(T)$ increases with higher frequencies, which is in agreement with the AC magnetic response of most spin glasses [31]. In fact, the $\chi''$ vs. T behavior indicates that all compounds of the $Dy_{3-x}Y_xTaO_7$ ($0 \leq x \leq 1$) solid solution are converging to a unique spin-glass state, since every curve seems to converge with the same tendency when the temperature is below $T_f$.

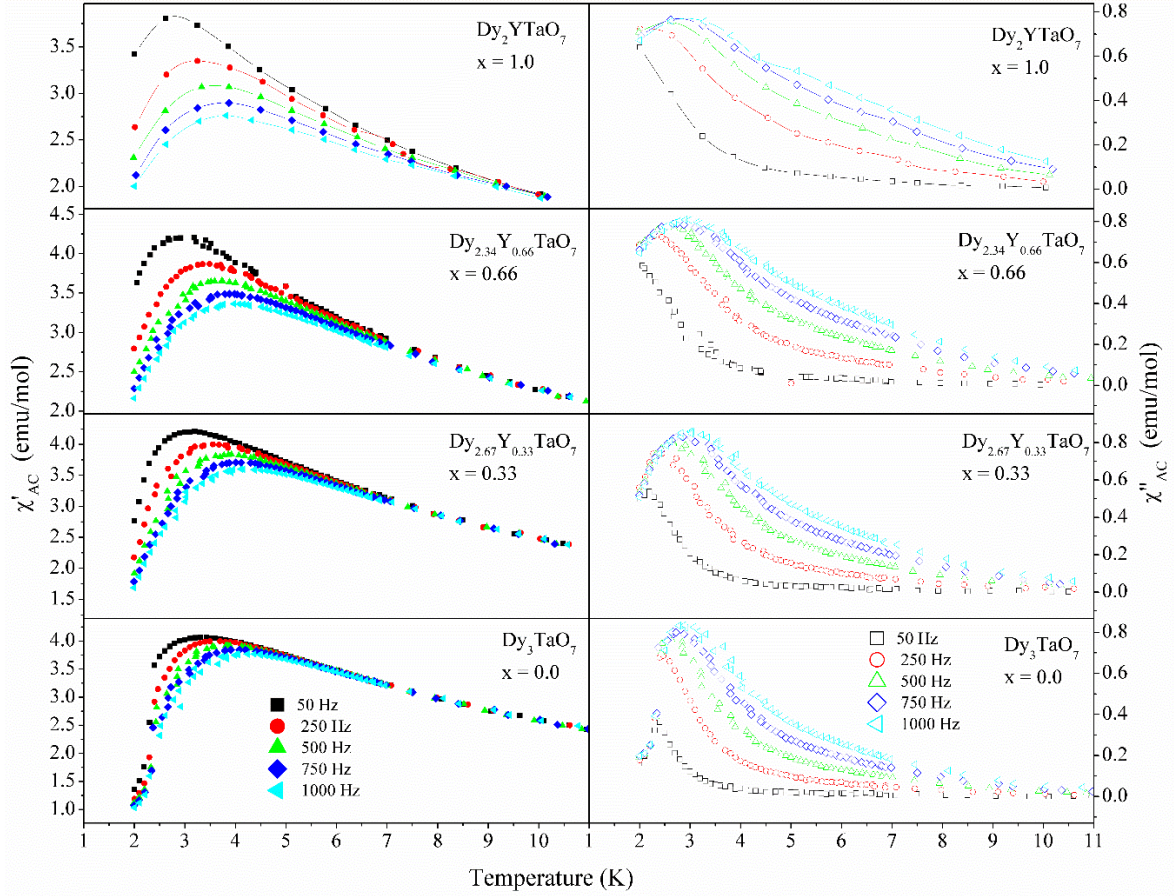

Fig. 6. (Color online) AC magnetic susceptibility measurements for samples of the $Dy_{3-x}Y_xTaO_7$ system. Left, in-phase ($\chi'$) component and right, out-of phase ($\chi''$) component at several frequencies at $\mu_0H_{ac}$ = 1 Oe.

The maximum change in freezing temperature, $K$, which identifies a canonical spin glass state can be estimated according to equation 2, where $\Delta T_f = T_f^{500\,Hz} - T_f^{50\,Hz}$ and $\Delta \log \omega = 1$, because this corresponds to one decade in frequency. The calculated K values are 0.14, 0.18, 0.15 and 0.2 for x = 0, 0.33, 0.66 and 1.0 respectively, and these are in good agreement with the value of K that ranges from 0.0045 to 0.28 for a canonical spin glass system [31].

$$K = \frac{\Delta T_f}{T_f \Delta(\log \omega)} \qquad \text{Equation 2}$$

The above equation only indicates whether the compound behaves or not as a spin glass, but it is not helpful to estimate the glassing temperature ($T_g$). On the other hand, the spin glass-like transition is better described by fitting the frequency dependence of $T_f$ to a critical power law (equation 3) [16,31,32] based on the theory of dynamical scaling analysis [32].

$$\tau = \tau_0 \left(1 - \frac{T_g}{T_f}\right)^{-zv} \qquad \text{Equation 3}$$

where $\tau = (2\pi f)^{-1}$, $\tau_o$ is the relaxation time of an individual particle or cluster moment, $T_g$ is the static glassy temperature, $z$ is the dynamical exponent, and $v$ is the critical exponent of the correlation length. The fitted parameters obtained through equation 3 are listed in Table VII.

*Table VII. Best fit values obtained from dynamical scaling analysis performed for x = 0, 0.33, 0.66, and 1.0 in the $Dy_{3-x}Y_xTaO_7$ system.*

| Sample | x value | $\tau_0$ (s) | $T_g$ (K) | $zv$ |
|---|---|---|---|---|
| $Dy_3TaO_7$ | 0 | $1.84 \times 10^{-5}$ | 3.19 | 1.56 |
| $Dy_{2.67}Y_{0.33}TaO_7$ | 0.33 | $7.38 \times 10^{-6}$ | 2.8 | 2.78 |
| $Dy_{2.34}Y_{0.66}TaO_7$ | 0.66 | $1.68 \times 10^{-6}$ | 2.21 | 5.8 |
| $Dy_2YTaO_7$ | 1.0 | $1.62 \times 10^{-7}$ | 1.02 | 23 |

As can be observed in Fig. 5, the $\chi'(T)$ curves display well-defined cusps that diminish in intensity, broaden and shift to higher temperatures with increasing frequency. The frequency of the cusp-temperature, $T_f$, obeys the equation 3. The $T_g$ values obtained from these fittings are consistent with those obtained from DC magnetic measurements ($T_{\chi\text{-max}}$, in Table V). A comparison between the DC and AC magnetic susceptibilities (see Table VI and VII) reveals that the temperature at which $\chi_{\text{-max}}$ occurs is similar to the $T_g$ estimated from fittings to Equation 3. Besides, the $T_g$ temperature coincides with that of the irreversibility temperature of $Dy_{2.34}Y_{0.66}TaO_7$ (see Fig. 3 B-2) (IT~2.25 K, $T_{\chi\text{-max}}$=2.3 K, $T_g$=2.21 K). It is worth mentioning here that we did not observe any $T_{\chi\text{-max}}$ for the x = 1.0 sample, but dynamical analysis predicts a $T_g$ of 1.02 K, which is lower than the lowest temperature reported in this work (1.8 K). Fig. 7 shows the typical linear variation of equation 3; all compounds fit well to this model and by comparing the $\tau_o$ values we can assume that the $Dy_{3-x}Y_xTaO_7$ system is a cluster spin glass, but the magnitude of its interactions become weaker as yttrium content increases in the lattice, this because the $\tau_o$ values slightly decrease.

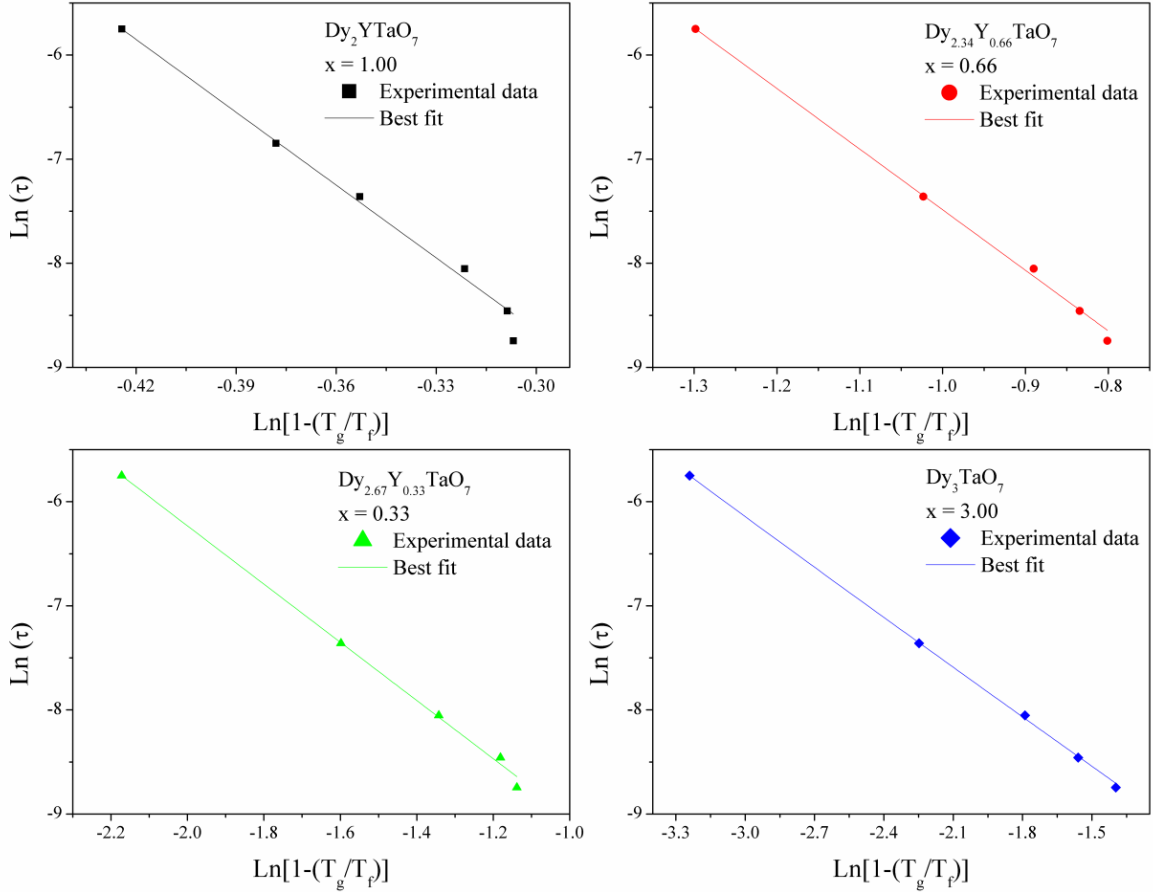

Fig. 7. (Color online) Linear variation of $\ln \tau$ vs. $\ln\left[1 - \left(T_g/T_f\right)\right]$ for the samples studied in this work. The values obtained in each fit are presented in Table VII.

On the basis of the DC and AC magnetic measurements, a spin glass behavior is clear in the $Dy_{3-x}Y_xTaO_7$ (0 ≤ x ≤ 1) system. Even more this system behaves as a canonical spin glass, but the obtained $\tau_o$ values are close to those observed for non-metallic systems [33], meanwhile the $z\nu$ values vary in a wide range and only two x-compositions agree well with the anticipated values for a cluster spin glass system [31,34]. For the x = 0 system, $Dy_3TaO_7$, the spin glass state can be directly associated with the $Dy^{3+}$-$Dy^{3+}$ distorted tetrahedra, leading to a quenched disorder with random J couplings. According to the Edwards-Anderson spin glass model [35], in this situation the system cannot satisfy all the couplings at the same time and becomes frustrated [36]. Following this idea, the spin glass state in $Dy_3TaO_7$ has as main component a quenched magnetic disorder. When $Y^{3+}$ substitutes $Dy^{3+}$ in the magnetic sublattice, an additional element appears and it can be associated to a chemical disorder, now with a nonmagnetic ion. It is hard to give an accurate explanation for the $T_g$ values dependence on x-composition, because these vary from 3 to 2 K. In this scenario, the possible antiferromagnetic order (see $\Theta_{CW}$ in Table VI) at the low temperature regime is strongly impeded from appearing because the system is frustrated (see $f_f$ values in Table VI). However, by comparing the decrement of $T_g$ values vs. the $Dy^{3+}$ content (see Table VII), it could be assumed that the $T_g$ change results from the solid solution chemical disorder, because in all compounds the structural distortions remain. The spin glass behavior could be reached as a consequence of the distorted $Dy^{3+}$ tetrahedra in the lattice that provide a huge number of J values into the magnetic lattice. In this way, the insertion of nonmagnetic yttrium ions into the crystal structure implies that the number of J couplings slightly diminishes, without breaking the spin glass state in the studied compositions.

## 4. Conclusions

We have successfully synthesized four compounds of the $Dy_{3-x}Y_xTaO_7$ solid solution: x = 0, 0.33, 0.66 and 1.0. Rietveld structural analysis showed a single crystal phase indexed in the *$C222_1$* space group. The linear change in the lattice cell parameters is consistent with the difference between the Dy and Y ionic radii.

Although $Dy_3TaO_7$ was previously reported to show antiferromagnetic order, we have shown that this compound behaves as a spin glass. In the same way, another compositions in the $Dy_{3-x}Y_xTaO_7$ system (x = 0.33, 0.66, and 1.0) also display spin glass behavior. Dynamic scaling analysis shows $T_g$ values to be close to those measured by DC magnetic susceptibility. The values of critical exponent and relaxation times suggest the existence of a cluster spin glass in $Dy_{3-x}Y_xTaO_7$.

**Acknowledgments**

The authors of this work are indebted to Dr. Francisco Morales, Dr. Pablo de la Mora, and Dr. E. Zeller for their advice, valuable assistance and useful suggestions to this work. We thank to Q.I. Cecilia Salcedo by her help regarding XRD measurements. This work was sponsored by PAPIIT (UNAM) IN-214313 project and the scholarship number 223355 granted by CONACyT México.